%% file: monopole-prl.tex
\documentclass[prl,superscriptaddress,floatfix,twocolumn,showpacs,preprintnumbers,amsmath,amssymb]{revtex4}

\usepackage{epsfig}
\usepackage{dcolumn}
\usepackage{bm}

\begin{document}

\title{ \boldmath Direct Search for Dirac Magnetic Monopoles in $p\bar{p}$
Collisions at $\sqrt{s} = 1.96~\rm{TeV}$\unboldmath}

\input{authors_plus.tex}


\begin{abstract}
We search for pair-produced Dirac magnetic monopoles in $35.7~\rm{pb}^{-1}$
of proton-antiproton collisions at $\sqrt{s} = 1.96~\rm{TeV}$ with the
Collider Detector at Fermilab (CDF).  We find no monopole candidates
corresponding to a $95\%$ confidence-level cross-section limit $\sigma <
0.2~{\rm pb}$ for a monopole with mass between $200$ and $700~{\rm
GeV}/c^2$.  Assuming a Drell-Yan pair production mechanism, we set a mass
limit $m > 360~{\rm GeV}/c^2$.
\end{abstract}


\maketitle

%
%

The existence of magnetic monopoles would add symmetry to Maxwell's
equations without breaking any known physical law.  More dramatically, it
would make charge quantization a consequence of angular momentum
quantization, as first shown by Dirac~\cite{DIRAC}.  With such appeal,
monopoles continue to excite interest and have been the subject of numerous
experimental searches.  

Grand unified theories predict monopole masses of about
$10^{17}~\rm{GeV/c^2}$, so cosmic ray experiments have searched
extensively for high-mass monopoles produced in the early universe.
Accelerator searches for low-mass monopoles have looked for the effects of
virtual monopole loops~\cite{MONOVIRTUAL, MONOVIRTUAL_L3, MONOVIRTUAL_D0},
but the results have been questioned~\cite{MONOVIRTUAL_PROBS}.  Detector
materials exposed to radiation from $p \bar{p}$ collisions at the Tevatron
have been examined for trapped monopoles, but the limit obtained depends on
the model for the trapping of monopoles in matter~\cite{MONOTRAPPED}.
Despite these efforts, magnetic monopoles have not been
discovered~\cite{MONOSUMMARY}.

Magnetic monopoles have magnetic charge $g$ satisfying the Dirac
quantization condition:
\begin{displaymath}
\frac{g e}{\hbar c} = \frac{n}{2} \Longleftrightarrow
\frac{g}{e} = \frac{n}{2\alpha}  \approx 68.5 \cdot n
\end{displaymath}
where $n$ is an integer and $\alpha$ is the fine structure constant. In this
search, we consider an $n=1$ monopole with mass less than $1~{\rm TeV}/c^2$,
spin $\frac{1}{2}$, and no hadronic interactions.  Monopoles are accelerated
by a magnetic field and are highly ionizing due to the large value of $g/e$.

This search uses a $35.7~{\rm pb}^{-1}$ sample of $p\overline{p}$ collisions
at $\sqrt{s} = 1.96~{\rm TeV}$ produced by the Fermilab Tevatron and
collected by the CDF II detector during 2003 using a special trigger.  The
detector consists of a magnetic spectrometer including silicon strip and
drift-chamber tracking detectors and a scintillator time-of-flight system,
surrounded by electromagnetic and hadronic calorimeters and muon
detectors~\cite{CDF}.  CDF uses a superconducting solenoid to produce a
$1.4~{\rm T}$ magnetic field.  The field is parallel to the beam direction,
which is taken as the $z$ direction, with $\phi$ the azimuthal angle, and
$r$ the radial distance in the transverse plane.  

The important detector components for this search are the central outer
tracker (COT)~\cite{COT} and the time-of-flight (TOF) detector~\cite{TOF},
both positioned inside the solenoid.  The coverage of the cylindrical COT
extends from a radius of $40~{\rm cm}$ to $137~{\rm cm}$ and to
psuedo-rapidity $|\eta| \sim 1$.  The COT consists of eight superlayers,
each containing 12 layers of sense wires.  The COT makes timing measurements
for track reconstruction as well as integrated charge measurements for
determining a particle's ionization energy loss $dE/dx$. The COT is
surrounded by 216 TOF scintillator bars, which run parallel to the beam line
and form a cylinder of radius $140~{\rm cm}$. Each TOF bar is instrumented
with a photomultiplier tube (PMT) on each end. The TOF measures both the
time and height of PMT pulses; the pulse height is typically used to correct
for discriminator-threshold time-slewing.  Due to their large ionization and
massive production of delta rays, monopoles in scintillator with velocity
$\beta > 0.2$ are expected to produce more than $500$ times the light from a
minimum-ionizing particle (MIP)~\cite{MONOSUMMARY, MACRO}.

%
%

We have built and commissioned a highly ionizing particle trigger that
requires large light pulses at both ends of a TOF scintillator bar.  The
trigger was designed to detect monopoles efficiently while consuming less
than $1~{\rm Hz}$ of the CDF data acquisition bandwidth.  The electronics
response of the TOF has been calibrated \cite{TOF_TRIGGER, THESIS_MIKE} to
account for non-linearities and channel-to-channel differences.  The trigger
thresholds of about $30~{\rm MIPs}$ are well below the expected response to
a monopole and have a negligible effect on the trigger efficiency.

In the CDF detector, a monopole is accelerated along the uniform
solenoidal magnetic field in a parabola slightly distorted by relativistic
effects.  Because no other particle mimics this behavior, the TOF acceptance
must be estimated from Monte Carlo simulation.  We have extended the GEANT
simulation~\cite{GEANT, GEANT_MONOPOLES, THESIS_PHIL, THESIS_MIKE} to handle
magnetic monopoles, including the acceleration from the magnetic field,
energy loss and multiple scattering~\cite{AHLEN}.

\begin{figure}[h]
\begin{center}
\epsfig{file=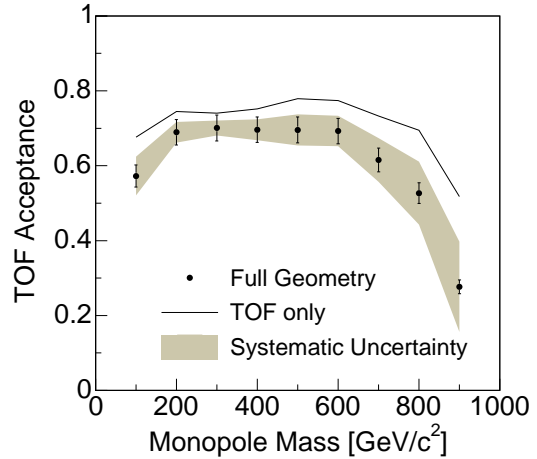,width=0.4\textwidth}
\caption{
  \label{fig:tofacc}
  The acceptance of the TOF for monopole pairs, as a function of monopole
  mass.  The band indicates the total systematic uncertainty.
}
\end{center}
\end{figure}

Because the monopole-photon coupling is large and non-perturbative, there is
no universally accepted field-theoretic calculation of magnetic-monopole
production.  However, monopole interactions with matter, such as scattering,
require only a replacement of the electric charge with the monopole's
effective charge $g \beta$.  This has led the authors of
Ref.~\cite{MONOTRAPPED} to adopt a heuristic production model by making the
same replacement for Drell-Yan monopole pair production, which we take as
our primary benchmark.

Either a monopole or anti-monopole must reach the TOF detector in order to
cause a trigger.  To calculate the TOF acceptance for the heuristic pair
production mechanism, we produce lepton Drell-Yan events with
Pythia~\cite{PYTHIA} with the lepton mass replaced by the monopole mass, and
weight events according to the additional velocity dependence.  The TOF
acceptance for monopole pairs simulated with GEANT is shown in Figure
\ref{fig:tofacc}. Light monopoles, accelerated strongly by the magnetic
field, tend to be swept out of the detector before reaching the TOF.  Heavy
monopoles, produced near threshold, suffer the same fate.

Because we are unable to test experimentally the model for material
interactions, we assign a systematic uncertainty of one half the total
calculated effect, measured by comparing the TOF acceptance for the full
simulation with a ficticious detector consisting of the TOF only.  The
material in the detector lowers the acceptance due to energy loss and
multiple scattering, a $3\%$ systematic error for intermediate-mass
monopoles. This method likely overestimates the uncertainty; varying the
energy-loss model between a naive model where $e \rightarrow g\beta$ and the
full treatment of Ref. \cite{AHLEN} has a negligible effect.

The TOF acceptance depends on the monopole production kinematics.  To
quantify this dependence, we consider separately the Drell-Yan mechanism
without the additional velocity dependence and with monopole production
uniform in the cosine of the polar angle in the center of mass frame. The
total variation in the acceptance is $10\%$.  We therefore present results
for our benchmark mechanism only, with the understanding that mass limits
for other production mechanisms can be inferred from the cross-section limit
with reasonable accuracy.

During each event, the TOF electronics makes a single measurement for each
PMT.  Light from other particles, called spoilers, can reach a PMT before
the light from monopoles, starting the charge integration.  If the monopole
light does not reach the PMT within the $20~{\rm ns}$ charge integration
window, the monopole's light will not be integrated and trigger will not
fire.  Our studies show that pure Monte Carlo underestimates the effect of
spoilers seen in data.  We therefore estimate the spoiler fraction by
embedding Monte Carlo produced monopoles in real $Z \rightarrow e^+e^-$
data.  Because these are high-mass central events produced by a Drell-Yan
mechanism, we expect the distribution of other particles in the event to be
similar to that of a monopole-pair production event.  We exclude the bars
with signals from the electrons and count the number of spoiler events,
which have real pulses arriving more than $20~{\rm ns}$ before the simulated
pulse from a magnetic monopole.

The systematic uncertainty is dominated by the uncertainty in the time
needed to integrate enough of the monopole's charge to cause a trigger.  To
quantify this effect, we note that rise times for TOF pulses are typically
less than $1~{\rm ns}$ and redo the calculation with a $15~{\rm ns}$
integration window.  We take one-half the difference as a systematic
uncertainty. Other effects, such as the dependence on luminosity, are much
smaller for our sample.  For a $400~{\rm GeV}/c^2$ monopole, the
spoiler fraction is $2\% \pm 1\%$ with a $3\%$ systematic uncertainty.

Massive monopoles can have low velocities causing them to arrive at the TOF
too late to cause a trigger.  The timing acceptance is calculated with a
Monte Carlo simulation by requiring pulses to arrive within the 54~{\rm ns}
timing window.  Only heavy monopoles move slowly enough to be affected: a
$900~{\rm GeV}/c^2$ monopole is out of time in $10\%$ of events.  This is a
negligible effect on lighter monopoles.

%
%

Monopoles curve in the $r z$ plane, in sharp contrast to electrically
charged particles, which curve in the $r \phi$ plane.  A specialized
reconstruction program isolates monopole candidates using data from the COT.
Candidates consist of coincident track segments composed entirely of hits
with large ionization, consistent with a straight line in the plane
perpendicular to the magnetic field.

The COT electronics encodes the integrated charge as the width of a hit,
which is the ionization measurement used for monopole candidate selection. A
typical MIP produces hit widths of about $20~{\rm ns}$.  An extrapolation of
the non-linear COT response for ordinary particles predicts that monopoles
would produce hit widths of about $230~{\rm ns}$ ($1000$ MIPs), still within
the dynamic range of the COT.  We do not use this extrapolation.  Instead we
cut in the tail of the width distribution from ordinary tracks, found to be
at $140~{\rm ns}$ ($50$ MIPs) in minimum-bias data collected with an open
trigger highly efficient for inelastic $p \bar{p}$ collisions.  Hits with
charge below this amount are not considered by the monopole reconstruction.
As magnetic monopoles have much greater ionization than the tracks used to
determine this cut, it has a negligible effect on the efficiency.

The default COT tracking algorithm first reconstructs track segments in each
of eight superlayers.  It checks for hits loosely consistent with a
straight-line, using a tolerance of $20~{\rm ns}$.  The identified hits in
each segment are then fit to a circular trajectory.  In the monopole
algorithm, the segments are required to be composed entirely of
high-ionization hits.  Also, because a monopole can be as slow as $\beta
\sim 0.1$ with changing transverse velocity, the usual timing assumption
($t_{{\rm flight}} = r / c$) cannot be used.  Instead, the time of flight to
each superlayer is varied between $r/c$ and $10 r/c$ in $5~{\rm ns}$
increments.

A monopole candidate consists of several $\phi$-coincident, low-curvature
segments.  From Monte Carlo simulation, we choose a loose cut on the segment
curvature $\rho < 0.001~{\rm cm}^{-1}$, which for an electron would
correspond to $p_{{\rm T}}> 4~{\rm GeV/c}$.  Likewise, the $\phi$ tolerance
is a loose $0.2~{\rm radians}$.  The remaining cuts are on the minimum
number of hits needed in a segment and on the total number of
$\phi$-coincident segments required for a monopole candidate.  By ignoring
the width cut, the segment-finding algorithm efficiency is measured in an
independent data sample using high-$p_{{\rm T}}$ tracks.  In this manner, we
choose a highly efficient cut requiring seven coincident superlayers with at
least eight hits in each segment.  This has a $94\%$ efficiency with a $1\%$
statistical uncertainty.  For these cuts, the efficiency for finding
high-mass monopole pairs calculated with the Monte Carlo simulation is
nearly $100\%$.  The efficiency for high-$p_{{\rm T}}$ electrons in
simulation, after removing the width cut, is also nearly $100\%$.  There are
real detector effects contributing a small inefficiency.

As an ionizing particle passes through matter, the most energetic electrons
form delta rays.  For highly relativistic low-mass monopoles, the
large number of delta rays confuses the segment finding algorithm, lowering
the efficiency.  We check that GEANT is properly producing delta-rays by
comparing the efficiency of monopoles to kinematically equivalent heavy-ions
simulated in the absence of a magnetic field.  We scale the efficiency
determined from Monte Carlo simulation to make the high-mass monopole
efficiency agree with the high-$p_{{\rm T}}$ track efficiency.  As the small
inefficiency from real detector effects cannot be measured directly on
monopoles, we take one half of the total inefficiency as a systematic
uncertainty: $3\%$ for $400~{\rm GeV}/c^2$ monopoles.  

\begin{figure}
\begin{center} 
\epsfig{file=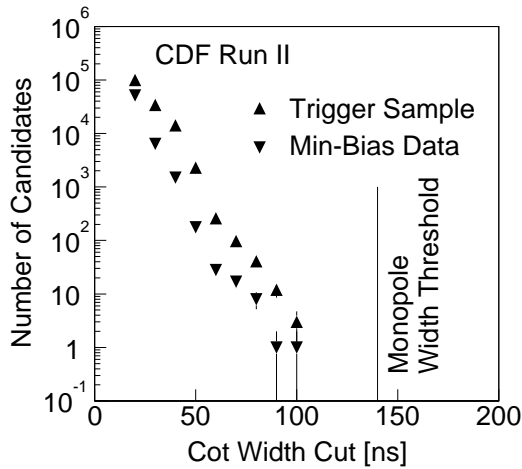,width=0.4\textwidth}
\caption{
 \label{fig:background} 
 Number of monopole candidates as a function of COT width cut in offline 
 reconstruction, in 100k minimum-bias events and the entire 130k event 
 trigger sample.  A 100 ns width cut corresponds to 15 MIPs, we expect
 monopoles to ionize more than 1000 MIPs or 232 ns.
}
\end{center}
\end{figure}

To estimate how effectively the monopole reconstruction rejects background,
we use minimum-bias data.  In $8 \times 10^5$ events, the event most like a
monopole has two coincident super-layers with seven hits per segment. Our
monopole requirements are much more stringent.  We require a seven-fold
coincidence of eight hits or more, resulting in extremely small background.
In the trigger sample the background is similarly small; the event most like
a monopole has two coincident super-layers with six hits per segment.  In
Fig. \ref{fig:background}, we count the number of monopole candidates passing
looser cuts on the hit width.

%
%

\begin{figure}
\begin{center} 
\epsfig{file=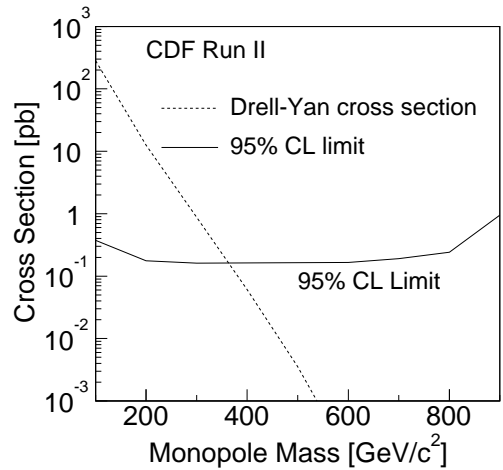,width=0.4\textwidth}
\caption{
  \label{fig:sensitivity}   
  The $95\%$ CL cross-section upper limit versus magnetic monopole mass.
  The theory curve for Drell-Yan monopole pair production intersects at the
  mass limit $m > 360~{\rm GeV}/c^2$.  }
\end{center}
\end{figure}

\begin{table}
  \begin{center}
    \begin{tabular}{| l | r |}
      \hline 
      Effect & Efficiency \\ 
      \hline
      TOF geometric (MC)   & $ 70\%  \pm 3\% \pm 3\%   $ \\
      TOF response         & $ 100\% $                   \\
      TOF spoilers         & $ 98\%  \pm 1\% \pm 3\%   $ \\
      TOF timing (MC)      & $ 99\%  \pm 1\% \pm 1\% $   \\
      \hline
      COT width cut        & $ 100\% $                   \\
      COT segment finding  & $  94\% \pm 1\% \pm 3\%   $ \\
      \hline
    \end{tabular}
    \caption{ 
    \label{tab:eff_sum} 
    Efficiency of the monopole search with statistical and systematic
    uncertainties for a monopole mass of $400~{\rm GeV}/c^2$. The full mass
    dependence is accounted for in the limit.  
    }
  \end{center}
\end{table}

None of the $130,000$ events from the monopole trigger sample passes the
candidate requirements, and we report a limit~\cite{THESIS_MIKE}.  Monopole
production limits are typically reported by the cross-section upper limit as
a function of monopole mass to minimize the dependence on a particular
production model. The expected number of events $N$ from a process with
cross section $\sigma$ and detector efficiency with acceptance $\epsilon$
after integrated luminosity $L$ is given by $N = L \epsilon \sigma$.  We
calculate the cross-section limit for zero observed events, based on the
efficiency summarized in Table \ref{tab:eff_sum} and a $6\%$ uncertainty in
the luminosity measurement~\cite{LUMI}.  We find the cross section for which
pseudo-experiments with efficiency and luminosity chosen randomly according
to their uncertainties yield one or more measured events $95\%$ of the time.

Our cross-section exclusion limit is shown in Figure \ref{fig:sensitivity}.
Our limit excludes monopole pair production for cross sections greater than
$0.2~{\rm pb}$~at the $95\%$ confidence level for monopole masses between
$200$ and $700~{\rm GeV}/c^2$.  For the Drell-Yan mechanism, this implies a
mass limit of $m > 360~{\rm GeV}/c^2$ at the $95\%$ confidence level.  This
is currently the best limit from a direct search.  Additional Run II data
will improve the sensitivity: another $300~{\rm pb^{-1}}$ extends the mass
reach by $100~{\rm GeV}/c^2$.

\input{acknowledge}

\bibliography{monopole-prl}

\end{document}

%% file: authors_plus.tex
%
%
%
\affiliation{Institute of Physics, Academia Sinica, Taipei, Taiwan 11529, Republic of China} 
\affiliation{Argonne National Laboratory, Argonne, Illinois 60439} 
\affiliation{Institut de Fisica d'Altes Energies, Universitat Autonoma de Barcelona, E-08193, Bellaterra (Barcelona), Spain} 
\affiliation{Baylor University, Waco, Texas  76798} 
\affiliation{Istituto Nazionale di Fisica Nucleare, University of Bologna, I-40127 Bologna, Italy} 
\affiliation{Brandeis University, Waltham, Massachusetts 02254} 
\affiliation{University of California, Davis, Davis, California  95616} 
\affiliation{University of California, Los Angeles, Los Angeles, California  90024} 
\affiliation{University of California, San Diego, La Jolla, California  92093} 
\affiliation{University of California, Santa Barbara, Santa Barbara, California 93106} 
\affiliation{Instituto de Fisica de Cantabria, CSIC-University of Cantabria, 39005 Santander, Spain} 
\affiliation{Carnegie Mellon University, Pittsburgh, PA  15213} 
\affiliation{Enrico Fermi Institute, University of Chicago, Chicago, Illinois 60637} 
\affiliation{Joint Institute for Nuclear Research, RU-141980 Dubna, Russia} 
\affiliation{Duke University, Durham, North Carolina  27708} 
\affiliation{Fermi National Accelerator Laboratory, Batavia, Illinois 60510} 
\affiliation{University of Florida, Gainesville, Florida  32611} 
\affiliation{Laboratori Nazionali di Frascati, Istituto Nazionale di Fisica Nucleare, I-00044 Frascati, Italy} 
\affiliation{University of Geneva, CH-1211 Geneva 4, Switzerland} 
\affiliation{Glasgow University, Glasgow G12 8QQ, United Kingdom} 
\affiliation{Harvard University, Cambridge, Massachusetts 02138} 
\affiliation{Division of High Energy Physics, Department of Physics, University of Helsinki and Helsinki Institute of Physics, FIN-00014, Helsinki, Finland} 
\affiliation{University of Illinois, Urbana, Illinois 61801} 
\affiliation{The Johns Hopkins University, Baltimore, Maryland 21218} 
\affiliation{Institut f\"{u}r Experimentelle Kernphysik, Universit\"{a}t Karlsruhe, 76128 Karlsruhe, Germany} 
\affiliation{High Energy Accelerator Research Organization (KEK), Tsukuba, Ibaraki 305, Japan} 
\affiliation{Center for High Energy Physics: Kyungpook National University, Taegu 702-701; Seoul National University, Seoul 151-742; and SungKyunKwan University, Suwon 440-746; Korea} 
\affiliation{Ernest Orlando Lawrence Berkeley National Laboratory, Berkeley, California 94720} 
\affiliation{University of Liverpool, Liverpool L69 7ZE, United Kingdom} 
\affiliation{University College London, London WC1E 6BT, United Kingdom} 
\affiliation{Massachusetts Institute of Technology, Cambridge, Massachusetts  02139} 
\affiliation{Institute of Particle Physics: McGill University, Montr\'{e}al, Canada H3A~2T8; and University of Toronto, Toronto, Canada M5S~1A7} 
\affiliation{University of Michigan, Ann Arbor, Michigan 48109} 
\affiliation{Michigan State University, East Lansing, Michigan  48824} 
\affiliation{Institution for Theoretical and Experimental Physics, ITEP, Moscow 117259, Russia} 
\affiliation{University of New Mexico, Albuquerque, New Mexico 87131} 
\affiliation{Northwestern University, Evanston, Illinois  60208} 
\affiliation{The Ohio State University, Columbus, Ohio  43210} 
\affiliation{Okayama University, Okayama 700-8530, Japan} 
\affiliation{Osaka City University, Osaka 588, Japan} 
\affiliation{University of Oxford, Oxford OX1 3RH, United Kingdom} 
\affiliation{University of Padova, Istituto Nazionale di Fisica Nucleare, Sezione di Padova-Trento, I-35131 Padova, Italy} 
\affiliation{University of Pennsylvania, Philadelphia, Pennsylvania 19104} 
\affiliation{Istituto Nazionale di Fisica Nucleare Pisa, Universities of Pisa, Siena and Scuola Normale Superiore, I-56127 Pisa, Italy} 
\affiliation{University of Pittsburgh, Pittsburgh, Pennsylvania 15260} 
\affiliation{Purdue University, West Lafayette, Indiana 47907} 
\affiliation{University of Rochester, Rochester, New York 14627} 
\affiliation{The Rockefeller University, New York, New York 10021} 
\affiliation{Istituto Nazionale di Fisica Nucleare, Sezione di Roma 1, University of Rome ``La Sapienza," I-00185 Roma, Italy} 
\affiliation{Rutgers University, Piscataway, New Jersey 08855} 
\affiliation{Texas A\&M University, College Station, Texas 77843} 
\affiliation{Istituto Nazionale di Fisica Nucleare, University of Trieste/\ Udine, Italy} 
\affiliation{University of Tsukuba, Tsukuba, Ibaraki 305, Japan} 
\affiliation{Tufts University, Medford, Massachusetts 02155} 
\affiliation{Waseda University, Tokyo 169, Japan} 
\affiliation{Wayne State University, Detroit, Michigan  48201} 
\affiliation{University of Wisconsin, Madison, Wisconsin 53706} 
\affiliation{Yale University, New Haven, Connecticut 06520} 
\author{A.~Abulencia}
\affiliation{University of Illinois, Urbana, Illinois 61801}
\author{D.~Acosta}
\affiliation{University of Florida, Gainesville, Florida  32611}
\author{J.~Adelman}
\affiliation{Enrico Fermi Institute, University of Chicago, Chicago, Illinois 60637}
\author{T.~Affolder}
\affiliation{University of California, Santa Barbara, Santa Barbara, California 93106}
\author{T.~Akimoto}
\affiliation{University of Tsukuba, Tsukuba, Ibaraki 305, Japan}
\author{M.G.~Albrow}
\affiliation{Fermi National Accelerator Laboratory, Batavia, Illinois 60510}
\author{D.~Ambrose}
\affiliation{Fermi National Accelerator Laboratory, Batavia, Illinois 60510}
\author{S.~Amerio}
\affiliation{University of Padova, Istituto Nazionale di Fisica Nucleare, Sezione di Padova-Trento, I-35131 Padova, Italy}
\author{D.~Amidei}
\affiliation{University of Michigan, Ann Arbor, Michigan 48109}
\author{A.~Anastassov}
\affiliation{Rutgers University, Piscataway, New Jersey 08855}
% ...
\collaboration{CDF Collaboration}
\noaffiliation

%% file: acknowledge.tex
We thank the Fermilab staff and the technical staffs of the participating
institutions for their vital contributions. This work was supported by the
U.S. Department of Energy and National Science Foundation; the Italian
Istituto Nazionale di Fisica Nucleare; the Ministry of Education, Culture,
Sports, Science and Technology of Japan; the Natural Sciences and
Engineering Research Council of Canada; the National Science Council of the
Republic of China; the Swiss National Science Foundation; the A.P. Sloan
Foundation; the Bundesministerium f\"ur Bildung und Forschung, Germany; the
Korean Science and Engineering Foundation and the Korean Research
Foundation; the Particle Physics and Astronomy Research Council and the
Royal Society, UK; the Russian Foundation for Basic Research; the Comisi\'on
Interministerial de Ciencia y Tecnolog\'{\i}a, Spain; in part by the
European Community's Human Potential Programme under contract
HPRN-CT-2002-00292; and the Academy of Finland.

%% file: monopole-prl.bbl
\begin{thebibliography}{19}
\expandafter\ifx\csname natexlab\endcsname\relax\def\natexlab#1{#1}\fi
\expandafter\ifx\csname bibnamefont\endcsname\relax
  \def\bibnamefont#1{#1}\fi
\expandafter\ifx\csname bibfnamefont\endcsname\relax
  \def\bibfnamefont#1{#1}\fi
\expandafter\ifx\csname citenamefont\endcsname\relax
  \def\citenamefont#1{#1}\fi
\expandafter\ifx\csname url\endcsname\relax
  \def\url#1{\texttt{#1}}\fi
\expandafter\ifx\csname urlprefix\endcsname\relax\def\urlprefix{URL }\fi
\providecommand{\bibinfo}[2]{#2}
\providecommand{\eprint}[2][]{\url{#2}}

\bibitem[{\citenamefont{Dirac}(1931)}]{DIRAC}
\bibinfo{author}{\bibfnamefont{P.~A.~M.} \bibnamefont{Dirac}},
  \bibinfo{journal}{Proc. Roy. Soc. Lond.} \textbf{\bibinfo{volume}{A133}},
  \bibinfo{pages}{60} (\bibinfo{year}{1931}).

\bibitem[{\citenamefont{Ginzburg and Schiller}(1998)}]{MONOVIRTUAL}
\bibinfo{author}{\bibfnamefont{I.~F.} \bibnamefont{Ginzburg}} \bibnamefont{and}
  \bibinfo{author}{\bibfnamefont{A.}~\bibnamefont{Schiller}},
  \bibinfo{journal}{Phys. Rev.} \textbf{\bibinfo{volume}{D57}},
  \bibinfo{pages}{6599} (\bibinfo{year}{1998}), \eprint{hep-ph/9802310}.

\bibitem[{\citenamefont{Acciarri et~al.}(1995)}]{MONOVIRTUAL_L3}
\bibinfo{author}{\bibfnamefont{M.}~\bibnamefont{Acciarri}} \bibnamefont{et~al.}
  (\bibinfo{collaboration}{L3}), \bibinfo{journal}{Phys. Lett.}
  \textbf{\bibinfo{volume}{B345}}, \bibinfo{pages}{609} (\bibinfo{year}{1995}).

\bibitem[{\citenamefont{Abbott et~al.}(1998)}]{MONOVIRTUAL_D0}
\bibinfo{author}{\bibfnamefont{B.}~\bibnamefont{Abbott}} \bibnamefont{et~al.}
  (\bibinfo{collaboration}{D0}), \bibinfo{journal}{Phys. Rev. Lett.}
  \textbf{\bibinfo{volume}{81}}, \bibinfo{pages}{524} (\bibinfo{year}{1998}),
  \eprint{hep-ex/9803023}.

\bibitem[{\citenamefont{Gamberg et~al.}(2000)\citenamefont{Gamberg,
  Kalbfleisch, and Milton}}]{MONOVIRTUAL_PROBS}
\bibinfo{author}{\bibfnamefont{L.~P.} \bibnamefont{Gamberg}},
  \bibinfo{author}{\bibfnamefont{G.~R.} \bibnamefont{Kalbfleisch}},
  \bibnamefont{and} \bibinfo{author}{\bibfnamefont{K.~A.}
  \bibnamefont{Milton}}, \bibinfo{journal}{Found. Phys.}
  \textbf{\bibinfo{volume}{30}}, \bibinfo{pages}{543} (\bibinfo{year}{2000}),
  \eprint{hep-ph/9906526}.

\bibitem[{\citenamefont{Kalbfleisch et~al.}(2004)\citenamefont{Kalbfleisch,
  Luo, Milton, Smith, and Strauss}}]{MONOTRAPPED}
\bibinfo{author}{\bibfnamefont{G.~R.} \bibnamefont{Kalbfleisch}},
  \bibinfo{author}{\bibfnamefont{W.}~\bibnamefont{Luo}},
  \bibinfo{author}{\bibfnamefont{K.~A.} \bibnamefont{Milton}},
  \bibinfo{author}{\bibfnamefont{E.~H.} \bibnamefont{Smith}}, \bibnamefont{and}
  \bibinfo{author}{\bibfnamefont{M.~G.} \bibnamefont{Strauss}},
  \bibinfo{journal}{Phys. Rev.} \textbf{\bibinfo{volume}{D69}},
  \bibinfo{pages}{052002} (\bibinfo{year}{2004}), \eprint{hep-ex/0306045}.

\bibitem[{\citenamefont{Giacomelli and Patrizii}(2003)}]{MONOSUMMARY}
\bibinfo{author}{\bibfnamefont{G.}~\bibnamefont{Giacomelli}} \bibnamefont{and}
  \bibinfo{author}{\bibfnamefont{L.}~\bibnamefont{Patrizii}}
  (\bibinfo{year}{2003}), \eprint{hep-ex/0302011}.

\bibitem[{\citenamefont{Blair et~al.}()}]{CDF}
\bibinfo{author}{\bibfnamefont{R.}~\bibnamefont{Blair}} \bibnamefont{et~al.},
  \emph{\bibinfo{title}{The \protect{CDF-II} {D}etector: {T}echnical {D}esign
  {R}eport}}, \eprint{FERMILAB-PUB-96-390-E}.

\bibitem[{\citenamefont{Affolder et~al.}(2004)}]{COT}
\bibinfo{author}{\bibfnamefont{T.}~\bibnamefont{Affolder}}
  \bibnamefont{et~al.}, \bibinfo{journal}{Nucl. Instrum. Meth.}
  \textbf{\bibinfo{volume}{A526}}, \bibinfo{pages}{249} (\bibinfo{year}{2004}).

\bibitem[{\citenamefont{Grozis et~al.}(2001)}]{TOF}
\bibinfo{author}{\bibfnamefont{C.}~\bibnamefont{Grozis}} \bibnamefont{et~al.},
  \bibinfo{journal}{Int. J. Mod. Phys.} \textbf{\bibinfo{volume}{A16S1C}},
  \bibinfo{pages}{1119} (\bibinfo{year}{2001}).

\bibitem[{\citenamefont{Ambrosio et~al.}(2002)}]{MACRO}
\bibinfo{author}{\bibfnamefont{M.}~\bibnamefont{Ambrosio}} \bibnamefont{et~al.}
  (\bibinfo{collaboration}{MACRO}), \bibinfo{journal}{Eur. Phys. J.}
  \textbf{\bibinfo{volume}{C25}}, \bibinfo{pages}{511} (\bibinfo{year}{2002}),
  \eprint{hep-ex/0207020}.

\bibitem[{\citenamefont{Bauer et~al.}()}]{TOF_TRIGGER}
\bibinfo{author}{\bibfnamefont{G.}~\bibnamefont{Bauer}} \bibnamefont{et~al.},
  \emph{\bibinfo{title}{The {T}ime-of-{F}light {T}rigger at {CDF}}},
  \bibinfo{note}{(submitted to NIMA)}, \eprint{FERMILAB-PUB-05-331-E}.

\bibitem[{\citenamefont{Mulhearn}(2004)}]{THESIS_MIKE}
\bibinfo{author}{\bibfnamefont{M.~J.} \bibnamefont{Mulhearn}}, Ph.D. thesis,
  \bibinfo{school}{Massachusetts Institute of Technology}
  (\bibinfo{year}{2004}).

\bibitem[{\citenamefont{Brun et~al.}()\citenamefont{Brun, Bruyant, Maire,
  McPherson, and Zanarini}}]{GEANT}
\bibinfo{author}{\bibfnamefont{R.}~\bibnamefont{Brun}},
  \bibinfo{author}{\bibfnamefont{F.}~\bibnamefont{Bruyant}},
  \bibinfo{author}{\bibfnamefont{M.}~\bibnamefont{Maire}},
  \bibinfo{author}{\bibfnamefont{A.~C.} \bibnamefont{McPherson}},
  \bibnamefont{and} \bibinfo{author}{\bibfnamefont{P.}~\bibnamefont{Zanarini}},
  \emph{\bibinfo{title}{\protect{GEANT3}}},
  \bibinfo{note}{\protect{CERN-DD/EE/84-1}}.

\bibitem[{\citenamefont{Bauer et~al.}(2005)}]{GEANT_MONOPOLES}
\bibinfo{author}{\bibfnamefont{G.}~\bibnamefont{Bauer}} \bibnamefont{et~al.},
  \bibinfo{journal}{Nucl. Instrum. Meth.} \textbf{\bibinfo{volume}{A545}},
  \bibinfo{pages}{503} (\bibinfo{year}{2005}),
  \eprint{doi:10.1016/j.nima.2005.01.326}.

\bibitem[{\citenamefont{Schieferdecker}(2001)}]{THESIS_PHIL}
\bibinfo{author}{\bibfnamefont{P.}~\bibnamefont{Schieferdecker}},
  \bibinfo{howpublished}{Undergraduate thesis, Ludwig Maximillian University}
  (\bibinfo{year}{2001}).

\bibitem[{\citenamefont{Ahlen}(1978)}]{AHLEN}
\bibinfo{author}{\bibfnamefont{S.~P.} \bibnamefont{Ahlen}},
  \bibinfo{journal}{Phys. Rev.} \textbf{\bibinfo{volume}{D17}},
  \bibinfo{pages}{229} (\bibinfo{year}{1978}).

\bibitem[{\citenamefont{Sjostrand et~al.}(2001)\citenamefont{Sjostrand,
  Lonnblad, and Mrenna}}]{PYTHIA}
\bibinfo{author}{\bibfnamefont{T.}~\bibnamefont{Sjostrand}},
  \bibinfo{author}{\bibfnamefont{L.}~\bibnamefont{Lonnblad}}, \bibnamefont{and}
  \bibinfo{author}{\bibfnamefont{S.}~\bibnamefont{Mrenna}}
  (\bibinfo{year}{2001}), \eprint{hep-ph/0108264}.

\bibitem[{\citenamefont{Klimenko et~al.}(2003)\citenamefont{Klimenko,
  Konigsberg, and Liss}}]{LUMI}
\bibinfo{author}{\bibfnamefont{S.}~\bibnamefont{Klimenko}},
  \bibinfo{author}{\bibfnamefont{J.}~\bibnamefont{Konigsberg}},
  \bibnamefont{and} \bibinfo{author}{\bibfnamefont{T.~M.} \bibnamefont{Liss}}
  (\bibinfo{year}{2003}), \eprint{FERMILAB-FN-0741}.

\end{thebibliography}
